\documentclass{article}
\usepackage{spconf,amsmath,graphicx}

\usepackage{pifont}
\newcommand{\cmark}{\ding{51}}%
\newcommand{\xmark}{\ding{55}}%

\usepackage{textcomp}
\usepackage{multirow}
\usepackage{array}
\usepackage[belowskip=-2pt,aboveskip=4pt]{caption}
\usepackage[justification=centering]{caption}

\title{Low Latency ASR for Simultaneous Speech Translation}
\makeatletter
\def\name#1{\gdef\@name{#1\\}}
\makeatother \name{{\em Thai Son Nguyen, Jan Niehues, Eunah Cho, Thanh-Le Ha}\\
	{\em Kevin Kilgour, Markus M\"uller, Matthias Sperber, Sebastian St\"uker, Alex Waibel}}

\address{Institute for Anthropomatics and Robotics Karlsruhe Institute of Technology\\
	{\small \tt firstname.lastname@kit.edu}
}
\begin{document}
%
\maketitle
\begin{abstract}
User studies have shown that reducing the latency of our simultaneous lecture translation system should be the most important goal. We therefore have worked on several techniques for reducing the latency for both components, the automatic speech recognition and the speech translation module. Since the commonly used commitment latency is not appropriate in our case of continuous stream decoding, we focused on word latency. We used it to analyze the performance of our current system and to identify opportunities for improvements. In order to minimize the latency we combined run-on decoding with a technique for identifying stable partial hypotheses when stream decoding and a protocol for dynamic output update that allows to revise the most recent parts of the transcription. This combination reduces the latency at word level, where the words are final and will never be updated again in the future, from 18.1s to 1.1s without sacrificing performance in terms of word error rate.
\end{abstract}

\section{Introduction}
In order for students to be able to follow a lecture by using our system's automatic lecture transcription and translation, the system's output needs to be as much in sync with the lecturer's speech and presentation as possible. Thus, the speech and translation components of the systems do not only need to run in real-time, but must produce output with as low a latency as possible. The high importance of a low latency is also the result of a user study and test that we conducted during real-world operation ~\cite{ip_mueller2016}. This paper addresses the problems of latency measurement and latency reduction for our speech transcription component. 


For applications that are turn based and operate on shorter queries, such as Google Voice and Apple Siri, the latency can be measured at utterance level, i.e. the response time after an utterances is finished. In these traditional applications users usually stop and wait for the results. But for our system, that is acting as an interpreter of a continuous, unsegmented stream of speech, the situation is different.
In our scenario there is no clear notion of utterance breaks in the speech, and thus measuring the latency becomes more difficult.
The traditional approach for latency measurements in real-time speech recognition systems found in literature uses either real-time factor or delay between the ending time of a segment, e.g. an utterance, and when the recognition result is available. Such measurements might capture the overall speed of the underlying recognition system but not the real latency as perceived by the users. To better capture the user experience, latency measures need to measure the delay as the word sequence is incrementally constructed---thus our focus on the word latency.
Further, it is not sufficient to look at an average of the overall latency of a whole test set. Instead we need to look at the variability of the latency and especially at the peak values in the per-word latency that are caused by problematic passages in the input audio. These peaks can lead to occasionally very high local latencies and need to be either avoided or dealt with appropriately.

In this paper we present two approaches to improve the latency of our lecture translator's speech recognition systems while maintaining its accuracy. The first approach uses a real-time recognition system, utilising an incremental decoding framework to decode continuous audio streams, in combination with a trace back of stable partial hypotheses. This approach is used to output whole portions, i.e. several words in sequence, of the final hypothesis as soon as possible. 

The second approach enhances the first one in combination with the display components by allowing to output partial hypotheses not only when they are stable, i.e., when it can be guaranteed that they will not change anymore in the future, but at any time as soon as they are available.
In unison with the display and translation components the recognition system is then allowed to correct itself later on, i.e. to revise the most recent history of its output, when a different word sequence has become more probable. 
Since very often the system will not need to correct itself, but the early output turns out to be the stable one, even though this could not have been guaranteed at the time it was passed on to the translation and display components, the latency of the system is reduced further this way.

Both approaches are shown to reduce the latency of the speech transcription component significantly from 18.1s to 1.1s without trade-off between latency and accuracy.

\section{Related work}
On-line speech recognition differs from off-line recognition in that latency is a crucial issue. Although latency in general refers to the response time of the recognition system, it has been defined in different ways in the past. For instance, in dialogue systems such as Google Voice~\cite{GoogleVoice}, latency is the time from when the user finishes speaking until the search results appear. In other related work on speech recognition for broadcast news~\cite{BroadcastNews}, latency measurement has included the time for the input to be completed. Note that carefully defining latency is important because only then we can optimize our recognition system in a systematic fashion.

A related but different concept, the \emph{real-time factor} (RTF), is calculated as the ratio between the utterance duration and its required decoding time. RTF is a common measure to evaluate the speed of a speech recognition system. Although distinct from the concept of latency, reducing the RTF can lead to a reduced latency in recognition systems, especially when the decoding starts after the input is completed. Recently, work to improve the latency of Apple's digital assistant Siri by boosting the pruning behaviour of a deep neural network (DNN) acoustic model~\cite{Pruning-DNN} resulted in a RTF reduction of 23\%. In these systems, there is usually a trade-off between the RTF and accuracy. Larger sizes of pruning beams, acoustic models or language models can lead to a better recognition accuracy at the cost of computing time.

Some recent papers on incremental speech recognition~\cite{Google-Incremental,Selfridge-Incremental} have addressed the latency problem in dialogue systems. The authors conducted a study about the stability which defines how much a word or hypothesis portion remains unchanged over the incremental decoding. Low latency is achieved by early putting out hypotheses when they reached a certain level of stability. However, no results have been published for latency measurements in real applications.

Unlike in dialogue systems, there are no markers for utterance boundaries in continuous recognition systems. To the best of our knowledge, there is not yet a standard approach for latency measurement for continuous recognition system in the literature. A study on broadcast news~\cite{BroadcastNews} used the delay between the ending time of a partial output and when the partial recognition is available. While a study on speech recognition in meetings~\cite{Meeting-Latency} measured the latency as the difference between the end time of a spoken word and the time when the word was output by the speech recogniser. 

\section{Latency definition}
\label{subsec::latency_definition}
In order to more precisely discuss different types of delays and compare them to the previous studies we define and distinguish: \textbf{Word latency} as the difference between the time a spoken word and the time when its transcription is available at the display component. \textbf{Commitment latency} is the difference between the end time of an audio segment (Section~\ref{sec:run-on_recognition}) or portion (Section~\ref{sec:stable_portion}) and when its transcription is available at the display component. \textbf{Word or commitment peak latency} is the highest word or commitment latency measured on a test set.
\vspace*{-0.2cm}
\section{Speech recognition system}

\subsection{Run-on recognition}
\label{sec:run-on_recognition}
Run-on recognition overlaps the decoding process with the audio recording process in order to reduce the latency. Our system used an adapted version of the run-on decoding described in \cite{Fugen-Thesis}. As in \cite{Fugen-Thesis} our system uses an audio segmenter in a separate process for pre-processing which writes the incoming audio stream into a shared memory with the decoder, while at the same time filtering out stretches of silence. We refer to everything between two stretches of silence as a \emph{segment}, which are usually several seconds long but could be as long as a lecture.

Our recognition system's search is re-initialised before processing a new segment and reads the segment's audio data from the shared memory while the audio segmenter continues to write to it. The audio data read in \emph{chunks} consisting of a fixed number of frames is incrementally decoded. The system therefore only has to wait until a chunk of audio is available in contrast to batch processing which requires complete segments before decoding.
\vspace*{-0.2cm}
\subsection{Stable hypothesis portion}
\label{sec:stable_portion}
The decoder tries the to find the most probable hypothesis. Normally, only at the end of a segment the most probable hypothesis is obtained. However, because waiting for an end-of-segment detected by the segmenter leads to a high latency, we use a partial trace-back \cite{Fugen-Thesis} for finding stable portions of the hypothesis early.
In our design, we detect partial hypotheses right after a chunk has been processed. Whenever a partial hypothesis is detected, its stable portion is extracted and delivered. The end of the portion will be tracked for the next detection.

\begin{figure}[t]
	\centering
	\includegraphics[width=\linewidth]{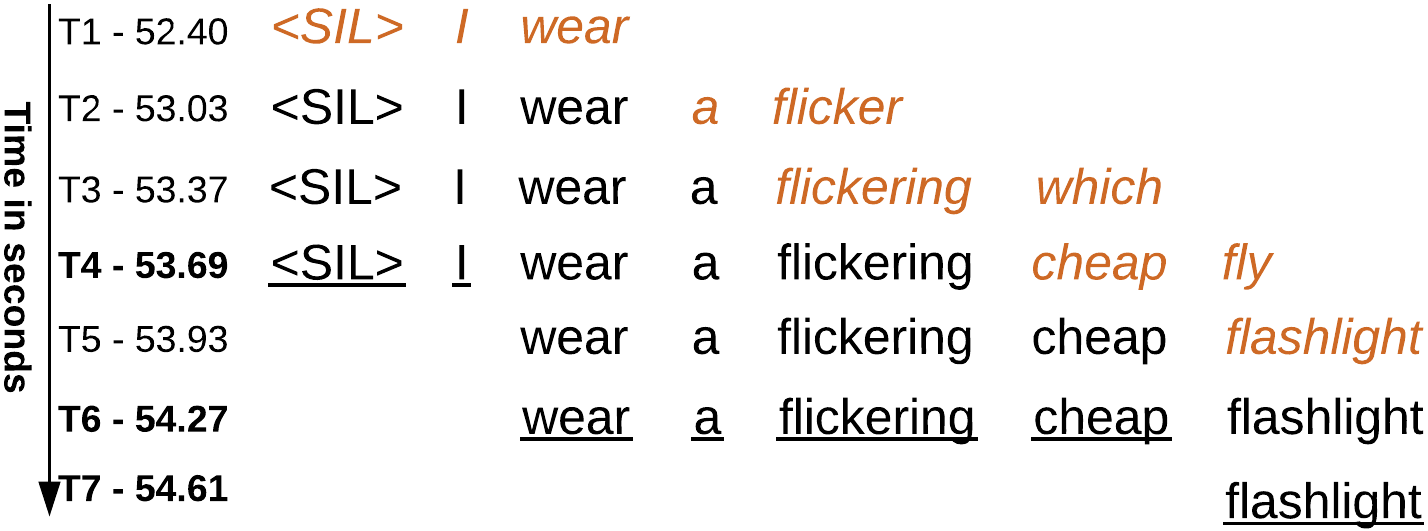}
	\caption{{\it An example of hypothesis update.}}
	\label{fig:hypothesis_update}
	\vspace*{-0.5cm}
\end{figure}
\vspace*{-0.2cm}
\subsection{Adaptive pruning}
\label{sec:adaptive_pruning}
Although our recognition system runs, on average, significantly faster than real time, we frequently encounter individual chunks which are processed much slower than real-time. This happens when encountering chunks that are difficult to decode, e.g. speech with background music or noises in which case the beam might fail to prune away competing paths effectively. This problem results in an unstable response time and introduces latency peaks. To overcome the problem, we use an adaptive pruning scheme. When an audio chunk is processed slower than real-time, we will narrow the beam to reduce the processing time of the following chunks. Once the recognition system has caught up again with the live audio the beam size is set back to its normal size.

\subsection{Hypothesis update}
\label{sec:hypothesis_update}
Next, we introduce another method that dramatically reduces the latency. We output probable parts of the unstable hypothesis and present them to the user. Later, the recognizer can revise its decision and overwrite the previous output if necessary. In this way, the recognition component does not need to wait until a stable portion or end-of-segment, instead it finds the most probable hypothesis every iteration of the incremental decoding, detects and sends the update portions to the display component.

Figure~\ref{fig:hypothesis_update} illustrates how this works in detail. In the example, the incremental decoding was performed 7 times and each time the most probable hypothesis was generated. The updated parts (italic text) were detected each time and sent to the display component. At T4, T6 and T7 the system detected the stable portions (underlined text). These had however already appeared as part of the unstable hypothesis at much earlier times. At T5 and T7, the hypotheses had new start times as described in Section~\ref{sec:stable_portion}.

Ignoring the words that are later replaced this algorithm can be seen as inducing a partition of stable hypothesis resulting in an improved latency without any accuracy loss. For example, only at T6 the system was sure about the stable hypothesis portion ``wear a flickering cheap'', but the parts of it were already sent, ``wear'' at T1, ``a'' at T2, ``flickering'' at T3 and ``cheap'' at T4. So the latency is again improved.

\subsection{Limiting the length of the partial hypothesis}
\label{sec:limit_length}
Since neither the segmenter nor the detection of partial hypothesis provide any guarantee regarding a maximal length of stable portions, we may still encounter situations in which the system does not output anything for a longer period of time. To deal with this issue, a threshold can be applied after which we force an output. If the waiting time exceeds this threshold, we simply output the most probable hypothesis at this point and track the end of the hypothesis.

\section{Experiments}

\subsection{System description}
We evaluate two baseline systems and three variants using the techniques described above for reducing latency. The first baseline which demonstrates batch processing, waits for completed segments before performing the whole decoding. The segments are generated by our integrated energy based segmenter. In the second baseline, we replace the batch processing with the run-on decoding described in Section~\ref{sec:run-on_recognition}. The decoded results are still produced for whole segments. Run-on decoding is employed in all three of the examined experimental systems.

The first experimental system, labeled \emph{Portion}, uses the algorithm from Section~\ref{sec:stable_portion} for finding stable hypothesis portions. The second variant, called \emph{ Update}, applies the update protocol explained in Section~\ref{sec:hypothesis_update}. Both of these utilise adaptive pruning. The third variant, named \emph{Update-NA}, applies the update protocol but without adaptive pruning. A chunk size of 40 frames is used in all run-on systems. Table~\ref{tab:system_summary} shows the summary of the applied techniques.

\begin{table}[t]
	\vspace{-0.0cm}	
	\centerline{
		\begin{tabular}{|c|cccc|}
			\hline
			\multicolumn{1}{|c|}{System} & Run-on & AP & PH & Update \\
			\hline \hline
			\emph{Baseline-1} & \xmark  & \xmark  & \xmark  & \xmark  \\
			\emph{Baseline-2} & \cmark & \cmark & \xmark  & \xmark  \\
			\emph{Portion}    & \cmark & \cmark & \cmark & \xmark  \\
			\emph{Update}     & \cmark & \cmark & \cmark & \cmark \\
			\emph{Update-NA}  & \cmark & \xmark  & \cmark & \cmark \\
			\hline
		\end{tabular}}
		\caption{\label{tab:system_summary} {\it System summary (AP = Adaptive Pruning, PH = Partial Hypothesis).}}
		\vspace{-0.2cm}
	\end{table}

\begin{table}[t]
	\setlength{\tabcolsep}{3pt}
	\centerline{
		\begin{tabular}{|c|cccc|}
			\hline
			System & WER & RTF & Commit. Latency & Word Latency \\
			\hline \hline
			Baseline-1 & 18.6 & 0.51 & 7.02 & 18.1 \\
			Baseline-2 & 18.4 & 0.68 & 0.92 & 10.2 \\
			Portion    & 18.5 & 0.68 & 1.72 & 2.10 \\
			Update     & 18.5 & 0.68 & 0.83 & 1.09 \\
			Update-NA  & 18.5 & 0.71 & 1.03 & 1.23 \\
			\hline
		\end{tabular}}
		\caption{\label{tab:overall_peformance} {\it Overall performance.}}
		\vspace{-0.5cm}
	\end{table}
	
All systems share the same basic setup. It uses a hybrid DNN/HMM acoustic model with log-Mel features. The acoustic model uses a context dependent phoneme setup with three states per polyphone. The DNN has an input window of +-7 frames, followed by 4 layers of 1,600 neurons and a classification output layer containing just over 8,000 neurons. We used a 4-gram language model with more than 150 thousand words for the decoding system.
\subsection{Performance evaluation \& Latency measurement}

The test data used for the evaluation includes 8 TED talks from the development set of the IWSLT 2015 evaluation campaign. First we evaluate the overall performance by averaging the measurements of WER, RTF, commitment latency and word latency over all talks. RTF is measured by the ratio between the processing time and the length of the processed audio segment. Commitment latency and word latency are measured as defined in Section~\ref{subsec::latency_definition}. Secondly, we measure and analyze the peaks in latency using both commitment and word latency.

All audio is fed into the ASR in real-time, as if it were being recorded by a microphone. The display component receives the sequence of uttered words and measures at when the words arrive. The RTF is measured directly by the ASR component, while latency is measured on the display side.

For the systems that send updated parts, we do not consider the latency of the intermediate words which are later replaced. Instead, we measure the latency to the time when a word has been updated the last time. We can then directly compare these measurements to the other systems.
\vspace{-0.1cm}

\section{Results}

\subsection{Overall performance}
Table~\ref{tab:overall_peformance} shows the performance of all systems on all test talks in terms of overall WER as well as average RTF, commitment latency and word latency.
	
\begin{figure}[t]
	\vspace{-0.3cm}
	\centering
	\includegraphics[width=\linewidth]{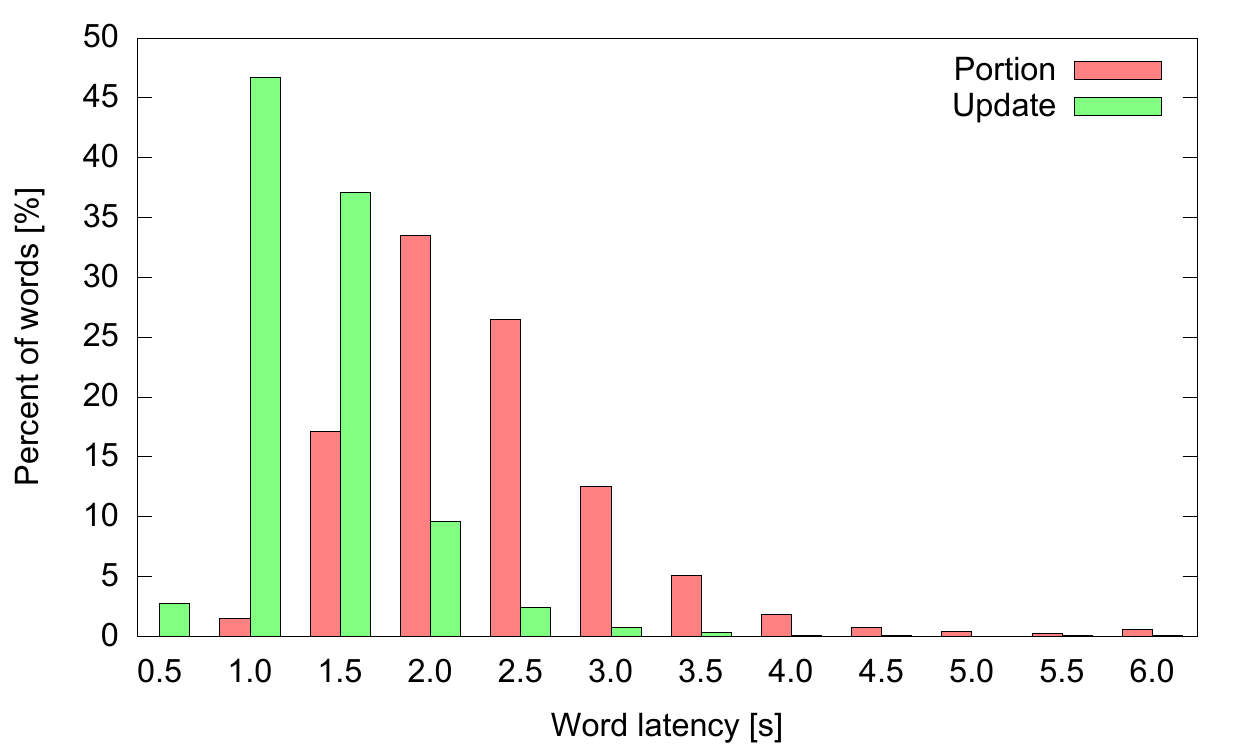}
	\caption{{\it Word latency distribution.}}
	\label{fig:word_latency}
	\vspace{-0.5cm}
\end{figure}
		
All the systems have similar WER performance. This confirms that our implemented algorithms did not change the accuracy. The batch processing \emph{Baseline-1} achieves a lower RTF than the other systems that employ run-on processing. This is because it is less efficient for the DNN acoustic model to process multiple smaller chunks than a few large chunks.

Despite its low RTF \emph{Baseline-1} has a large commitment latency since in the batch processing this latency mostly reflects the processing time of the segments. \emph{Portion} has a larger commitment latency than \emph{Baseline-2} and \emph{Update} since it needs to wait until the output can be guaranteed to be stable. \emph{Baseline-2} demonstrates that we can significantly reduce the commitment latency by following the run-on design. Note also that commitment latency, word latency, and RTF are only loosely correlated, indicating that commitment latency and RTF are not sufficient for evaluating the latency of the continuous recognition systems, and justifying our introduction of word latency.

\emph{Baseline-2} is especially interesting in this regard, because it has a low commitment latency but a very high word latency. This demonstrates the need for committing recognition results as quickly as possible in order to achieve a low latency. In this sense, \emph{Portion} and \emph{Update} perform better than the others.

As a more detailed analysis, we provide the statistics in Figure~\ref{fig:word_latency}. It shows the latency distribution of all uttered words in the test set. We only focus on \emph{Portion} and \emph{Update}. According to the diagram, most spoken words are recognised within 2 seconds in \emph{Update}, and 3.5 seconds in \emph{Portion}.
\subsection{Peak latency}
Looking only at the overall latency, the difference in latency between \emph{Update} and \emph{Update-NA} appears small. However, in practice we noticed that \emph{Update-NA} has a much higher larger peak latency. Table~\ref{tab:peak_latency} shows the peak latency of all systems. \emph{Portion} required up to 23 seconds to identify a stable hypothesis while the worst word for \emph{Update} was displayed after only 9 seconds. \emph{Update-NA} has a similar overall latency as \emph{Update}, but its peak latency is much worse. The results of \emph{Baseline-2} emphasises the need for word latency measurements.
\begin{table}[t]

\centerline{
	\begin{tabular}{|c|cc|}
		\hline
		System & Max Commit. Latency & Max Word Latency \\
		\hline \hline
		Baseline-1 & 93.4 & 230  \\
		Baseline-2 & 2.7  & 145  \\
		Portion    & 21.8 & 23.0 \\
		Update     & 2.7  & 9.0  \\
		Update-NA  & 19.3 & 23.5 \\
		\hline
	\end{tabular}}
	\caption{\label{tab:peak_latency} {\it Peak latency.}}
	\vspace{-0.2cm}
\end{table}

\begin{figure}[t]
	\vspace{-0.2cm}
	\centering
	\includegraphics[width=0.9\linewidth]{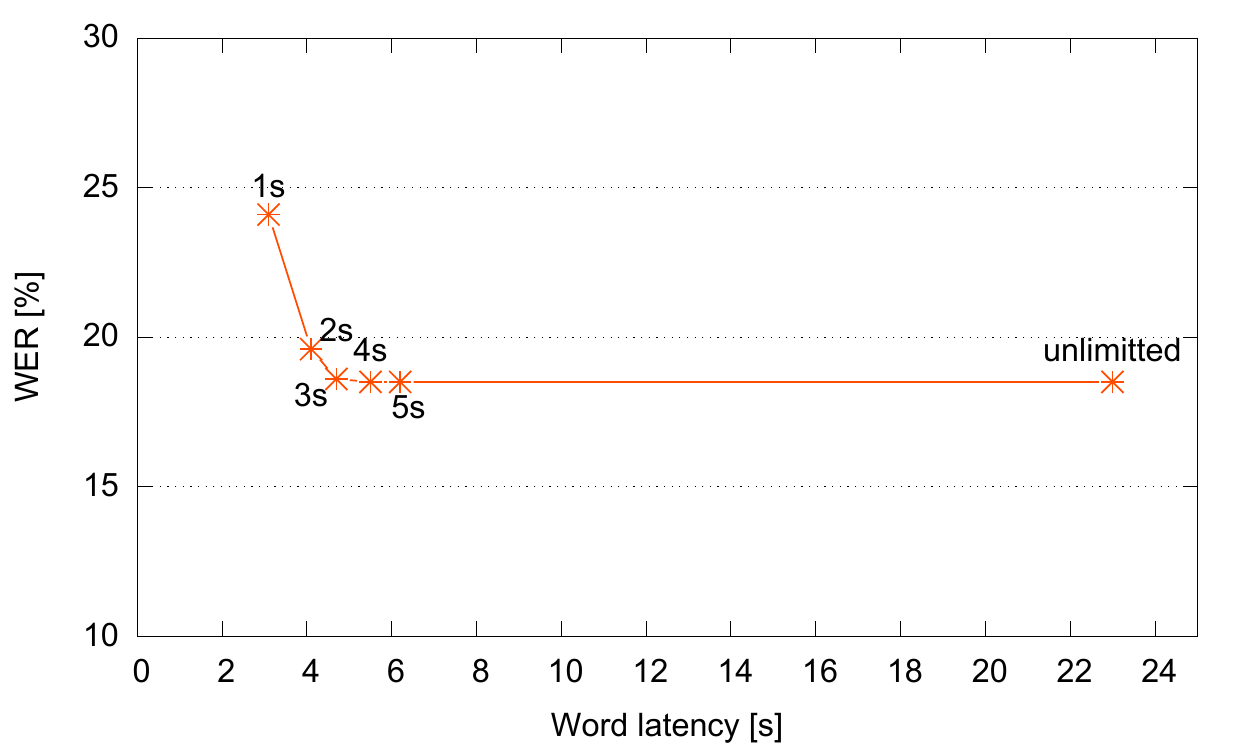}
	\caption{{\it WER vs. Peak Latency.}}
	\label{fig:peak_latency}
	\vspace{-0.5cm}
\end{figure}

To explore whether the peak latency can be further improved, we performed an experiment by applying length limitings techniques from Section~\ref{sec:limit_length} to \emph{Portion}. Figure~\ref{fig:peak_latency} presents the accuracy and the peak latency of the system at   different maximal threshold settings from unlimited over 5~s down to 1~s. We can see that the system imposing a 3s threshold hardly impacts its accuracy.

\section{Conclusion}
We have presented an evaluation for exploring and analysing different problems of latency and latency measurement in our continuous speech recognition system. We have also presented several techniques to deal with these latency problems. The latency improvement not only enhances the usability of our lecture translation system, but also enables the transcriptions to be in sync with the slides and gestures of the lecturer.

\bibliographystyle{IEEEbib}
\bibliography{refs}

\end{document}